\documentclass[runningheads]{llncs}
\usepackage[T1]{fontenc}
\usepackage{graphicx}
\begin{document}
%

\title{From Waterfallish Aerospace Certification \\
onto Agile Certifiable Iterations}

%
%

\author{J. Eduardo Ferreira Ribeiro\inst{1}\orcidID{0000-0002-1894-3993} \and
Mário Zenha-Rela\inst{2}\orcidID{0000-0003-1985-9344} \and
João Gabriel Silva\inst{2}\orcidID{0000-0002-4800-5201}}

\authorrunning{J. Eduardo Ferreira Ribeiro et al.}
%
\institute{Department of Informatics Engineering, Faculty of Engineering, University of Porto, Porto, 4200-465, Portugal \url{https://www.fe.up.pt} \and CISUC, Department of Informatics Engineering, Coimbra, 3030-290, Portugal \url{https://www.uc.pt/fctuc/dei/}}
\maketitle              
\begin{abstract}
Agile software development is becoming increasingly popular in the aerospace industry because of its capability to accommodate requirement changes. However, safety-critical domains require compliance with strict regulations such as the \emph{DO-178C} avionics standard, which demands thorough documentation. The main challenge of this constraint is not the content itself, but rather the comprehensive traceability from system-level requirements to all sorts of testing and verification evidence, including who did what, when, and to which artifact. Currently, this is mostly a manual activity performed at the end of the project, which blocks efforts to agilize the development of software for aerospace applications.
In this paper, we present a strategy and tools that support the generation of \textit{continuous documentation} complying with \emph{DO-178C} requirements. By iteratively creating the \emph{DO-178C} documentation associated with each software component and seamlessly merging it with the previously generated documentation, we open the way to truly \textit{continuous certifiable} iterations, an evolution from the current Waterfallish industry practice. The proposed mechanisms and tools were co-designed and validated with aerospace industry professionals, thereby confirming its applicability and usefulness. The generated artifacts show that document automation is feasible in the aerospace industry, opening the way for more widespread adoption of Agile practices in this highly regulated~sector.

\end{abstract}
\section{Introduction}
\label{Introduction}
Safety-critical systems are heavily regulated because of the severe impact of failures, including death, injury, property damage, and environmental harm~\cite{Ribeiro2023}. Therefore, companies in safety-critical domains must comply with specific standards to ensure confidence in the quality and safety of these systems and to provide the evidence necessary for certification~\cite{Ribeiro2023}. Traditionally, safety-critical systems engineering has relied on the Waterfall model to meet these standards, which includes the generation of safety-related documentation for regulatory certification~\cite{Ribeiro2023,RE_ribeiro2023beyond}. However, standards such as \emph{DO-178C} do not dictate a specific development model~\cite{RE_DO178C}, allowing flexibility in the adoption of other methods as long as adequate evidence demonstrates that applicable safety-level objectives are met~\cite{Ribeiro2023,RE_ribeiro2023beyond,RE_DO178C,RE_Rierson2013}. Meanwhile, Agile methods have emerged in various domains to enhance software development management, respond quickly to customer needs, and introduce shorter feedback loops, thus reducing the time to market~\cite{Ribeiro2023,RE_Marques2013}. Des\-pi\-te these benefits, one of the concerns raised in safety-critical environments for adopting Agile methods concerns how documentation is managed and interconnected due to its heavy dependence on manual processes~\cite{Ribeiro2023,RE_SilvaCardoso2022}.
In safety-critical applications, documentation is crucial to ensure the required quality and safety (e.g., specifications, analyses, reviews, and testing)~\cite{tordrup_meshing_2020}. Therefore, extensive documentation and traceability are mandated by certification authorities. Since manual generation of such documentation is still a common practice in the aerospace industry, this overwhelming effort, mostly due to stringent traceability needs, is blocking the widespread adoption of Agile practices in this highly regulated industry.

In this paper, we present an innovative strategy and tools to automate document management that streamlines the creation of certifiable documentation, from requirement specification to verification and validation (V\&V). It enables \textit{continuous certification}, i.e. the iterative generation of certification documents that mirrors the iterative development of each software component.

This approach builds on our previous works ~\cite{Ribeiro2023,RE_ribeiro2023beyond,RE_SilvaCardoso2022} using the Scrum framework~\cite{RE_Schwaber2020}. Designed explicitly for safety-critical aerospace development, our proposed automation mechanisms help organizations efficiently produce and reuse documents from previous incremental deliveries, while complying with the \emph{DO-178C} standard.


This study involved industry professionals from an aerospace software subcontractor and access to IBM DOORS\footnote{https://www.ibm.com/docs/en/ermd/9.7.0?topic=overview-doors}, an industry-specific tool compliant with the DO-330~\cite{RE_DO330} standard for aerospace software development. The co-design and iterative validation with industry professionals with more than 20 years of industry experience ensured that our proposed mechanisms for automating documentation are applicable in real-world aerospace scenarios.

This study was driven by the following research questions:

\begin{itemize}
    \item \textbf{RQ1}: Can automation mechanisms for documentation-related processes enhance the software certification process?
    \item \textbf{RQ2}: Which strategies can optimize a Document Management Tool (DMT) to make it \emph{quasi-autonomous}?
    \item \textbf{RQ3}: Is it feasible for a DMT to generate certifiable documentation automatically?
\end{itemize}

Section~\ref{Relatedwork} presents an overview of related work, including the benefits and challenges of automation, case studies and successful implementations. Section~\ref{FunctionalityStreamsModelProcess} briefly introduces the Scrum4DO178C, a software development process model that we designed and is currently being validated in a project with the major European aerospace manufacturer~\cite{Ribeiro2023,RE_ribeiro2023beyond,RE_SilvaCardoso2022}. It adopts a Scrum framework~\cite{RE_Schwaber2020} and servez as the testbed for the development of our strategy for automating documentation processing, management, and reuse while complying with \emph{DO-178C}. In Section~\ref{MechanismsforAutomating}, the core of the paper, we detail our proposal for improving efficiency and effectiveness in certification through automation. Section~\ref{ThreatstoValidity} discusses the main threats to validity identified during the study. Finally, Section~\ref{Conclusion} concludes the paper by revisiting the research questions, summarizing its main findings, and highlighting areas for further research.

\section{Related Work}
\label{Relatedwork}

The impact and overheads of documentation in the aerospace industry are well known. The study presented in ~\cite{youn_software_2015} highlighted the strict requirements for documentation throughout the software life cycle as a contributing factor to the high cost of commercial aviation systems. This is because of the extensive demands of the \emph{DO-178C} standard, which is a set of guidelines for the software development process in the aviation industry. This demand is fully justified: according to~\cite{hilderman_understanding_2014}, the \emph{DO-178C} standard provides several benefits, two of which are directly related to documentation. First, the precise documentation of parameter data items, unused software structures, and high-level requirements (HLR) and low-level requirements (LLR) leads to more thorough testing. This documentation ensures that all the aspects of the software are tested and verified, resulting in greater safety and reliability. Second, the \emph{DO-178C} standard's rigorous and consistent documentation, modularization, and enforcement of documented modern engineering principles and reviews improve the software's reusability, and the author states that reusability is a crucial aspect of software development. The \emph{DO-178C} standard significantly improves this aspect by ensuring that all techniques and processes mentioned above are followed. Moreover,~\cite{kennedy} emphasized the complexity and importance of \emph{DO-178C} standard-compliant certification and its inherent dependency on documentation. This standard requires relevant knowledge and experience within each organization and poses a challenge for start-ups in the aviation industry~\cite{dmitriev_lean_2020}.

Therefore, exploring software projects' documentation automation was central to our work, and a comprehensive survey of the existing literature and tools on the subject was essential to identify opportunities for improvement. In the following we refer to some of the studies most relevant to our work. In ~\cite{ahmadi_achachlouei_document_2021} Jupyter Notebook\footnote{https://jupyter.org/} appears to be a potential candidate for a documentation system because of its open-source nature and the use of JavaScript Object Notation (JSON) as the notebook document format. The Jupyter Notebook is a web-based interactive computational environment for creating and sharing documents containing live code, equations, visualizations, and narrative text. It has gained popularity in data science and scientific computing owing to its ease of use and ability to combine code, text, and visualizations in a single document. Jupyter was a useful inspiration, despite the fact that our problem that does not require interactive editing, since all we need is the capability to generate content in a pre-specified format (a template) and fill it with imported content from the development environment, which is a git-based project repository with tasks managed by Jira. \cite{comoretto_documentation_2020} mentions DocSteady\footnote{https://github.com/lsst-dm/docsteady}, a tool that generates LaTeX test documents by extracting information from Jira\footnote{https://www.atlassian.com/software/jira} using a model-based systems engineering approach.  This allows for reduced time to produce V\&V documents, better integration with the project's software system engineering model, and full system requirements traceability. We adopted a similar approach, but targeted the entire development lifecycle. Moreover, in the aerospace domain, IBM DOORS is the \textit{de facto} requirements management tool; therefore, our solution was intimately tied to DOORS.  Another study~\cite{lee_introduction_2012} discussed the use of Doxygen\footnote{https://www.doxygen.nl/} as a useful tool for automatically generating documents from the source code for flight software. Doxygen significantly reduces the time and complexity of the development process by allowing the generation of documents directly from source code. We also adopted Doxygen to grab tagged contents directly from the source files so that developers could embed directives directly from the Integrated Development Environment (IDE). In~\cite{xsdoc}, the authors proposed using XSDoc, a set of tools that can support the documentation process for source code written in Java and C++ programming languages, and models described in the Unified Modeling Language (UML). XSDoc can be easily integrated with IDEs, as we did with Doxygen. The authors argue, and we we fully agree, that the editing, revision, and evolution of the documentation contents are facilitated, controlled and structured, by integrating documentation into the software development environment. In our context, IBM DOORS is agnostic to the programming language used and supports traceability as long as hooks to the project files are provided. 


In summary, the aerospace software community is well aware of the challenges of complying with certification requirements. Several existing tools provide valuable insights into how to automate the software documentation process, but none target compliance with industry standards, namely the \emph{DO-178C} as we did. Such a specialized tool requires context, a reference development process that provides a specific workflow to fit in, and from where its requirements can be extracted. This will be the focus of the next section.

\section{Scrum4DO178C Process}
\label{FunctionalityStreamsModelProcess}

\begin{figure}[bt]
    \centerline{\includegraphics[width=0.65\textwidth]{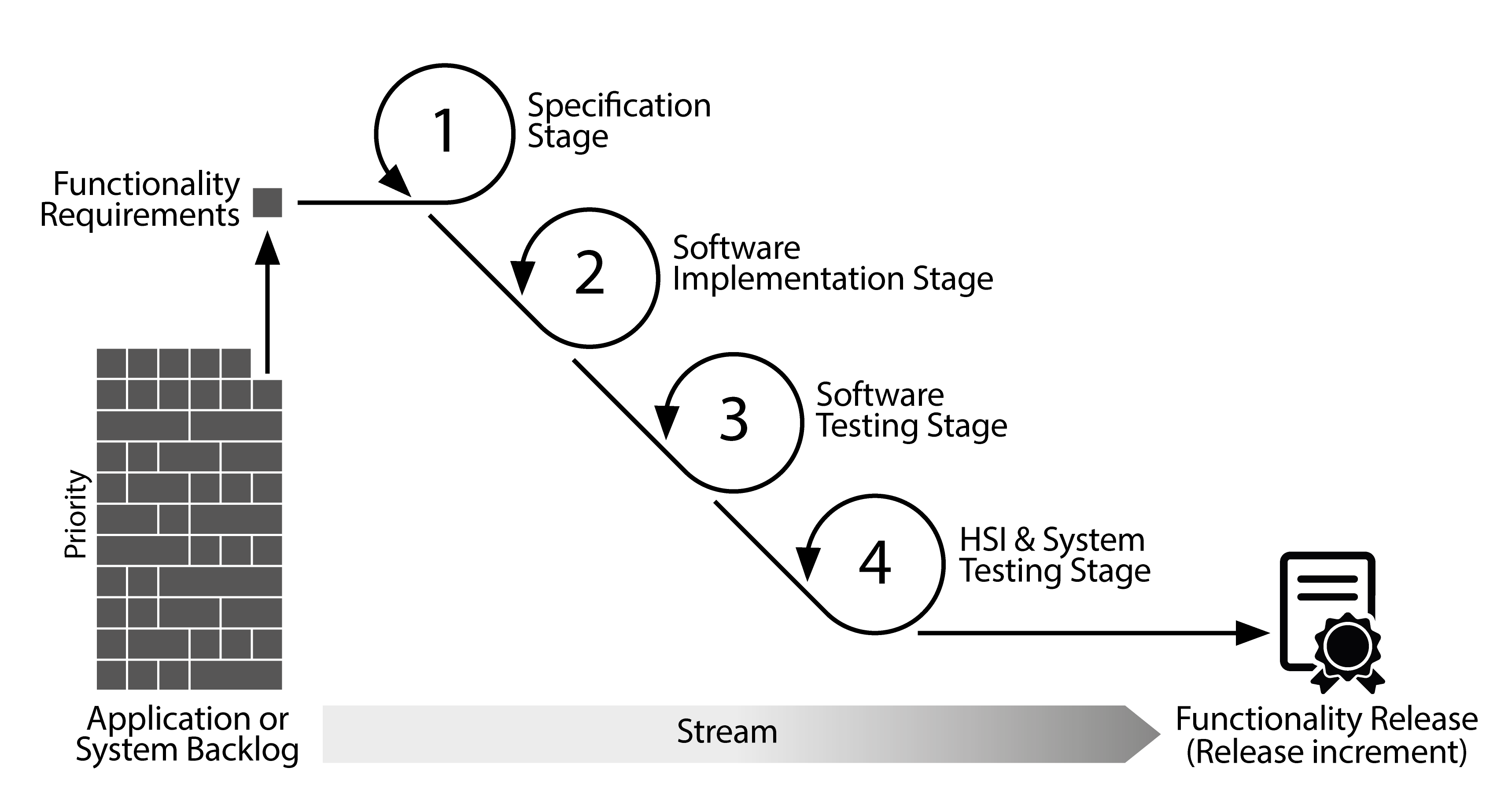}}
    \caption{Flow chart showing functionality stages.}
    \label{FStagesFlow}
\end{figure}

The \textit{Agile-Friendly Aerospace Software Development Process}(Scrum4DO178C)~\cite{Ribeiro2023} is a structured approach designed specifically for Agile safety-critical aerospace software development that complies with the \emph{DO-178C} standard. This process utilizes the Scrum framework (Type-A and Type-C)~\cite{RE_Schwaber2020,Sutherland2011}. It aims to reduce the timeline length between the V-Model phases, 
by improving the management of the requirements. 
The Scrum4DO178C is divided into four stages, as shown in Figure~\ref{FStagesFlow}. Each stage takes place in a sequence, with the work of each stage isolated from the others, following Scrum Type-A. Several stages are performed in parallel, following Scrum Type-C. The four stages of the process are: i) Specification, ii) Software Implementation, iii) Software Testing, iv) Hardware \& Software Integration (HSI) and System Testing.

To ensure compliance with the \emph{DO-178C} standard, specific tasks and sub-tasks were defined for each developed functionality in each stage. These tasks are designed to meet all levels of V\&V required by the standards and produce the necessary documentation and evidence to support certification. At the end of each stream, a certifiable output is delivered, which then undergoes an additional final V\&V of the release (Hardening the Product Backlog Item (PBI)\footnote{PBI is a smaller requirement increment of work that can be completed during a specific established timeframe~\cite{RE_Schwaber2020}.} or sprint\footnote{Sprint is a fixed-length iteration of one month or less, creating a consistent sequence for the team~\cite{RE_Schwaber2020}.}) as a final step.



By breaking down requirements into smaller, manageable PBIs and developing and testing them in iterative cycles (sprints) the Scrum4DO178C process actually breaks one single large V into a sequence of 'micro-Vs' around a set of related PBIs. Therefore, the Scrum4DO178C model serves as the Agile foundation for the development of mechanisms for automating the generation of  \emph{DO-178C} documentation, as described below.

\section{Automating \emph{DO-178C} Documentation}
\label{MechanismsforAutomating}

The prescriptive nature of \emph{DO-178C} can be used advantageously to automate the creation of documents required for certification. This standard imposes a comprehensive set of tasks and documents that must be delivered. Actually, this is a tree of tasks and related documents, starting with system specification, instantiated as HLRs, and detailed into LLRs down to the software components, unit tests, and their integration into higher-level components up to system testing. Every step needs to be fully documented with comprehensive tracking of artifacts and personnel\footnote{ Different safety levels (A to E) require different levels of checks.}. Thus, in simple terms, certification documentation is an exhaustive report of every activity and related artifacts for each requirement. This list must be complemented by a set of comprehensive traceability matrices covering every artifact (e.g., test results), its location in the project repository, timestamps and the individual(s) responsible. 
While most of the documentation content is done manually (requirements, architecture, design, coding, tests, and V\&V activities), the traceability matrices don't need to. These matrices are actually large spreadsheet files. Creating and manually checking those files is a daunting task, as it requires inputs from various tools such as project monitoring and requirements management tools (e.g., Jira and IBM DOORS), CI/CD pipelines, and versioning tools (e.g,. Bitbucket\footnote{https://bitbucket.org/product}), among others.

Our strategy to automate the generation of \emph{DO-178C} documentation was as follows:
\begin{enumerate}
    \item Use a template for the certification document(s) and input its content from relevant project repository files. Because the standard is very specific about the contents needed (e.g., requirements, review reports), a simple configuration file provides this mapping.
    \item Grab new items from project files using predefined Doxygen tags.
    \item Input any new requirements into IBM DOORS to ensure full tracking.
    \item Track every change to the project files at pull requests and update the traceability matrices. Identification of anomalous file changes.
    \item At specific events (end of sprint or release) generate a new 'potentially certifiable package' with the required certification documents. 
\end{enumerate}

Note how we mirror the Agile nature of software development in the certification process: the documentation is created as iteration deliverables always synchronized with the related PBIs of the sprint or release. Our solution currently addresses the Software Requirement Document (SRD)\footnote{The SRD outlines software system specifications, detailing required functionality, performance, interfaces, and system constraints. It ensures clear, traceable, and verifiable software requirements crucial for safety and regulatory compliance in airborne systems.~\cite{Ribeiro2023,RE_ribeiro2023beyond}.} and the Software Configuration Index (SCI)\footnote{The SCI identifies the software configuration and can contain a set of configuration items, such as the level of control and management applied to software configuration items throughout the development lifecycle~\cite{Ribeiro2023,RE_ribeiro2023beyond}.}. 

Two complementary tools were developed to support this strategy, the Certification Manager Tool (CMT) and the Certification Packager Tool (CPT). CMT monitors the work environment for changes during pull requests. It's goal is to collect all the information related to a specific requirement being developed, from the initial PBI to its full integration in the codebase, after all stages have been successfully passed. It grabs the content of a specific header (.h) file from a webhook triggered by a pull request event in Bitbucket. This header file contains new requirements that must be imported into IBM DOORS so that each requirement is mapped to the ID requirement generated by DOORS. This tool also generates a Software Design Artifact (.docx document) containing information on these requirements and commits it to a specific branch in Bitbucket, along with the new header file. 

At the end of each sprint or release, on request, the CPT is responsible for gathering all the information collected by the CMT and producing a potentially certifiable SCI Report ready to be delivered for certification. It retrieves from Bitbucket all the binary files related to Jira tickets specified in a configuration file as input and saves them in a specific folder ('data-package') inside the project repository.


Having outlined the overall strategy, in the following subsections, we describe the four Scrum4DO178C stages 
to better clarify where we are connecting the tools into this \emph{DO-178C} compliant process. 

\subsection{Specification Stage}
\begin{figure}[bt]
    \centerline{\includegraphics[width=\textwidth]{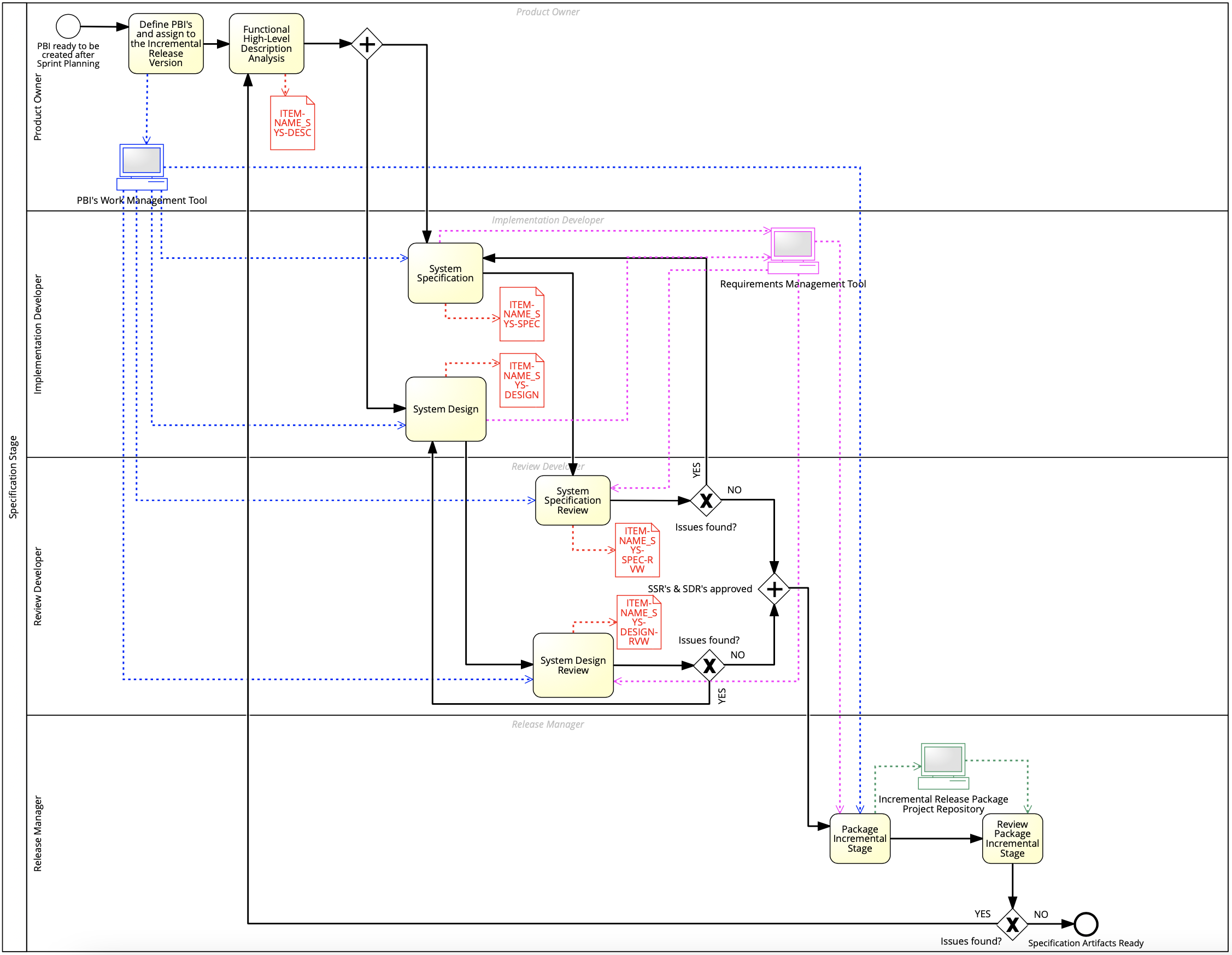}}
    \caption{Overview of the Specification Stage.}
    \label{bpmn-specification-stream}
\end{figure}

The system specifications and designs are described and reviewed during this stage (Fig.~\ref{bpmn-specification-stream}). The CMT automatically tracks all changes to the files.

The Specification Stage involves several key activities, including:
\begin{itemize}
    \item \textbf{Defining PBIs and assigning them to the Incremental Release Version}: this step entails defining PBIs related to the Specification Stage in the Work Management Tool (WMT), such as Jira, and associating each PBI with a specific release.
    \item \textbf{Conducting a Functional High-Level Description Analysis}: this involves a technical description of previously defined functionalities.
    \item \textbf{Developing a System Specification}: create a system requirements document stored in the Requirements Management Tool (IBM DOORS).
    \item \textbf{Conducting a System Specification Review}: a checklist is generated to ensure that all essential aspects of the document were addressed.
    \item \textbf{Creating a System Design}: this step involves elaborating on a design requirements document, which is also stored in IBM DOORS.
    \item \textbf{Conducting a System Design Review}: review the design requirements and generate a checklist to ensure that all crucial aspects have been addressed.
\end{itemize}
Except for defining PBIs, a work organization activity, the above tasks are mandated by \emph{DO-178C}. Note how prescriptive it is and how objectively the outputs (files) can be identified. This simplifies the automated traceability. The last two steps are Scrum4DO178C specific:
\begin{itemize}
    \item \textbf{Generating an Incremental Package}: A package containing documentation from the Specification Stage is assembled and placed in the project repository. This step is fully automated using CMT.
    \item \textbf{Reviewing the Incremental Package}: This step involves manually reviewing the package to ensure that all the necessary files are included.
\end{itemize}

The key strategy here is to track changes in all artifacts that are the outputs of each activity. This tracking involves not only the artifact but also its authors and sequence. Thus, this is actually about following the mandatory workflow and tracking file changes, as they are git-managed artifacts. At the end of this stage, a new Data Package was assembled, incorporating all certification artifacts generated thus far. It is important to emphasize that this package only contains the log of events and artifacts; thus, it is not yet in a format adequate for certification.

\subsection{Software Implementation Stage}
\begin{figure}[bt]
    \centerline{\includegraphics[width=\textwidth]{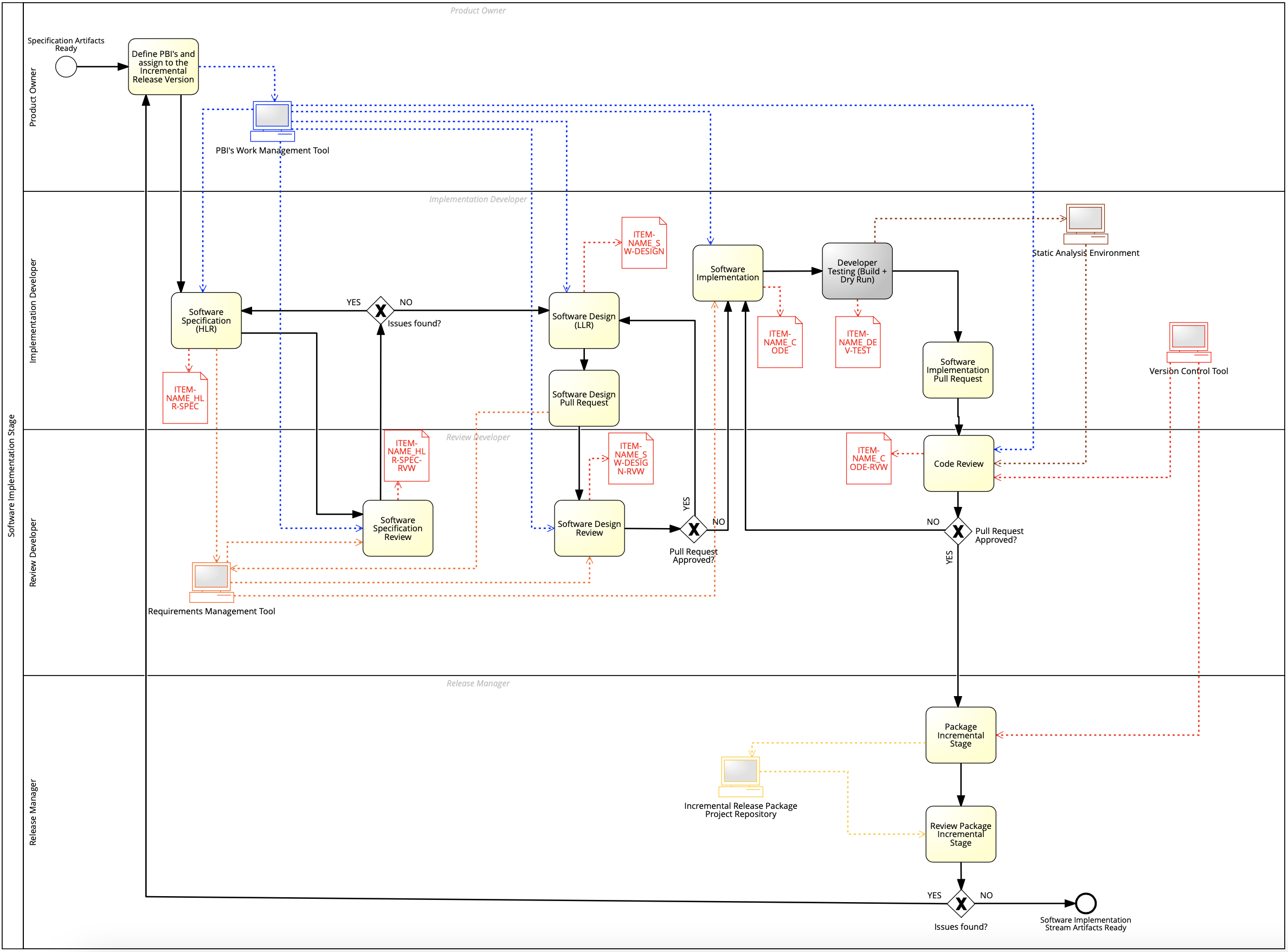}}
    \caption{Overview of the Software Implementation Stage.}
    \label{bpmn-software-implementation-stream}
\end{figure}

The Software Implementation Stage (Fig.~\ref{bpmn-software-implementation-stream}) marks the beginning of software construction and involves the definition of high- and low-level requirement specifications (HLRs and LLRs), reviewing them, and implementing the software accordingly. Again, the CMT keeps tracking all files changed.

This Stage encompasses various key activities, such as:
\begin{itemize}
    \item \textbf{Defining Product Backlog Items (PBIs) and assigning them to the Incremental Release Version}: PBIs related to the current stage are defined using the previous stage's HLRs as inputs.
    \item \textbf{Elaborating on Software Specification (HLR)}: involves elaborating on HLRs, that are directly entered into the RMT (IBM DOORS).
    \item \textbf{Conducting a Software Specification Review}: review and generate a checklist to ensure that all essential aspects are addressed.
    \item \textbf{Elaborating on Software Design (LLRs)}: create LLRs in the IDE and submit them to a Code Review in BitBucket.
    \item \textbf{Creating a Software Design Pull Request}: to trigger a Review.
    \item \textbf{Conducting a Software Design Review}: review the LLRs and generate a  checklist to ensure that all crucial aspects have been addressed.
    \item \textbf{Implementing the Software}: this step involves coding the functionalities according to the LLRs.
    \item \textbf{Conducting Developer Testing (Build + Dry Run)}: test the build and run tests locally on the developer's machine.
    \item \textbf{Creating a Software Implementation Pull Request}: this step involves creating a pull request in the VCS (bitbucket) to trigger Code Review.
    \item \textbf{Conducting the Code Review}: review and generation of a checklist to ensure that all essential coding aspects have been addressed.
\end{itemize}
Again, the above are tasks mandated by \emph{DO-178C} while the last two steps are Scrum4DO178C specific:
\begin{itemize}
    \item \textbf{Generating an Incremental Package}: A package containing documentation from the Software Implementation Stage \textit{and earlier stages} is assembled and stored in the project repository. This step is automated using the Configuration Management Tool (CMT).
    \item \textbf{Conducting a Review of the Incremental Package}: This step involves manually reviewing the package to ensure that all necessary files have been correctly generated.
\end{itemize}

Note that the Configuration Management Tool is actually creating and tracking on the fly the traceability matrix; when a PBI leads to a software specification (HLR), a new entry must be created and tracked, along with all dependent file changes. 

\subsection{Software Testing Stage}
The Unit and Integration Testing Stage is one of the most critical factors, as the focus of \emph{DO-178C} is software quality. Unit and integration tests are developed and run in Dry Run mode i.e. execute the tests before integration in the main branch. Typically, these tests are run on the developer's local machine but may also need to run on a dedicated environment. If no issues are found during the Dry Run, a new Data Package  containing all the artifacts produced thus far can be assembled at the end of this stage. 

\begin{figure}[bt]
    \centerline{\includegraphics[width=\textwidth]{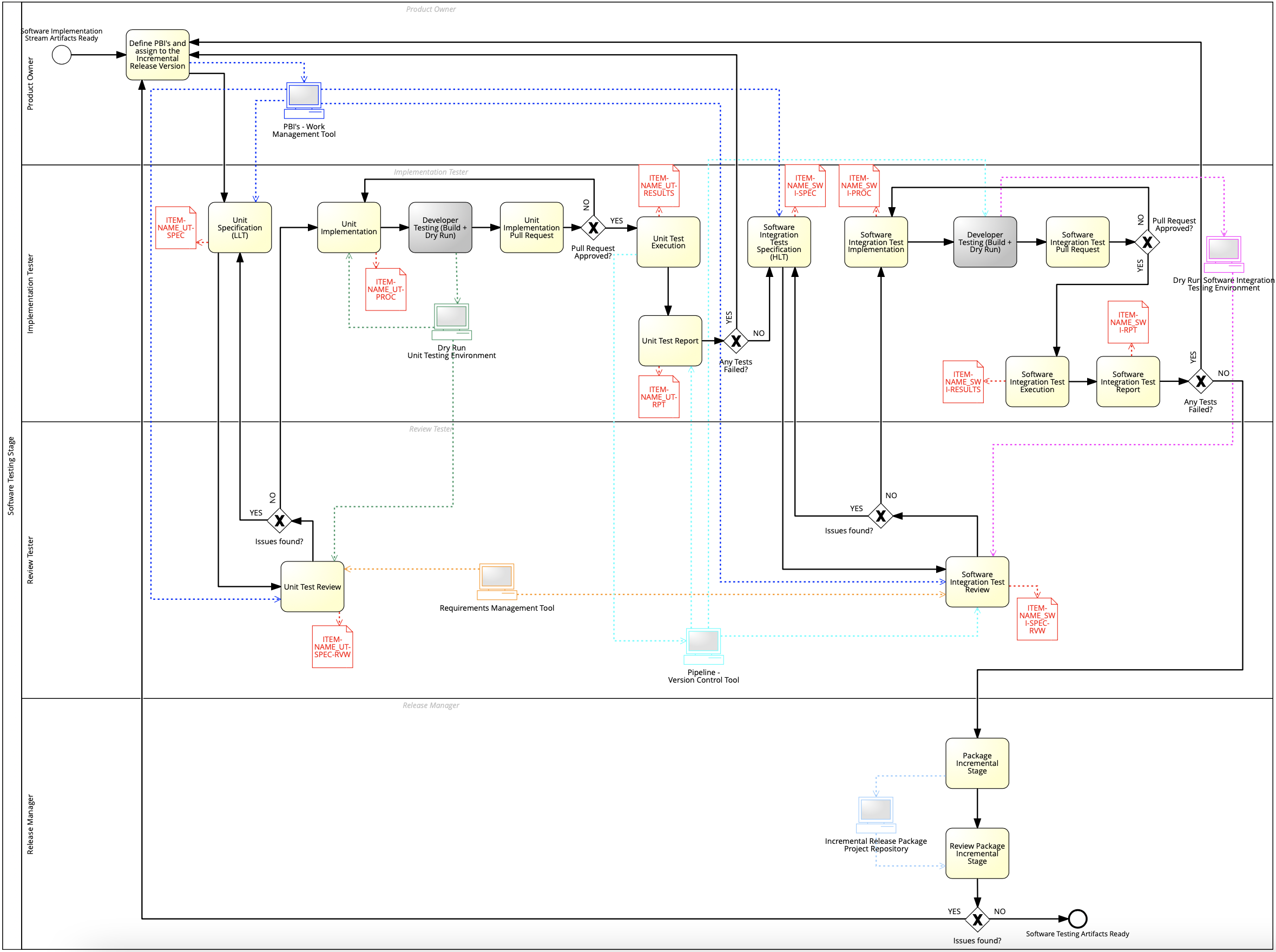}}
    \caption{Overview of the Software Testing Stage.}
    \label{bpmn-software-testing-stream}
\end{figure}

Several key activities are involved in the Software Testing Stage, namely:
\begin{itemize}
    \item \textbf{Defining PBIs and assigning them to the Incremental Release Version}: define new PBIs and associating each PBI with a specific release. 
    \item \textbf{Elaborating on Unit Test Specification}: this step involves detailing the low-level tests (LLTs).
    \item \textbf{Conducting a Unit Test Review}: review LLTs and generate a checklist to ensure that all essential aspects of the tests have been addressed.
    \item \textbf{Implementing the Unit Tests}: coding the LLTs.
    \item \textbf{Conducting Developer Testing (Build + Dry Run)}: test the build and run tests locally on the developer's machine.
    \item \textbf{Creating a Unit Implementation Pull Request}: trigger remote exec.
    \item \textbf{Conducting Unit Test Execution}: this step involves executing all the developed LLTs in an environment distinct from the local developer machine.
    \item \textbf{Elaborating the Unit Test Report}: results of the LLTs execution.
    \item \textbf{Elaborating the Software Integration Tests Specification known as high-level tests (HLT)}.
    \item \textbf{Conducting a Software Integration Test Review}: review and generate a checklist to ensure that all essential aspects have been addressed.
    \item \textbf{Implementing the Software Integration Tests}: this step involves coding the HLTs that integrate specific LLRs.
    \item \textbf{Conducting Developer Testing (Build + Dry Run)}: build and run tests locally on the developer's machine.
    \item \textbf{Creating a Software Integration Test Pull Request}: to trigger the remote Software Integration Test Execution.
    \item \textbf{Conducting Software Integration Test Execution}: this step involves executing all the developed HLTs in a dedicated environment.
    \item \textbf{Elaborating on Software Integration Test Report}: results of HLTs.
\end{itemize}
And the last two steps, Scrum4DO178C specific:
\begin{itemize}
    \item \textbf{Generating an Incremental Package}: A package containing documentation from the Software Testing Stage and \textit{previous stages} is assembled by the CMT and placed in the project's repository. 
    \item \textbf{Conducting a Review of the Incremental Package}: manually reviewing the package to ensure that all necessary files have been included.
\end{itemize}

\subsection{HSI \& System Testing Stage}
The final stage of the process involves performing HSI (Hardware \& Software Integration) 
under conditions that mimic production environments.

\begin{figure}[bt]
    \centerline{\includegraphics[width=\textwidth]{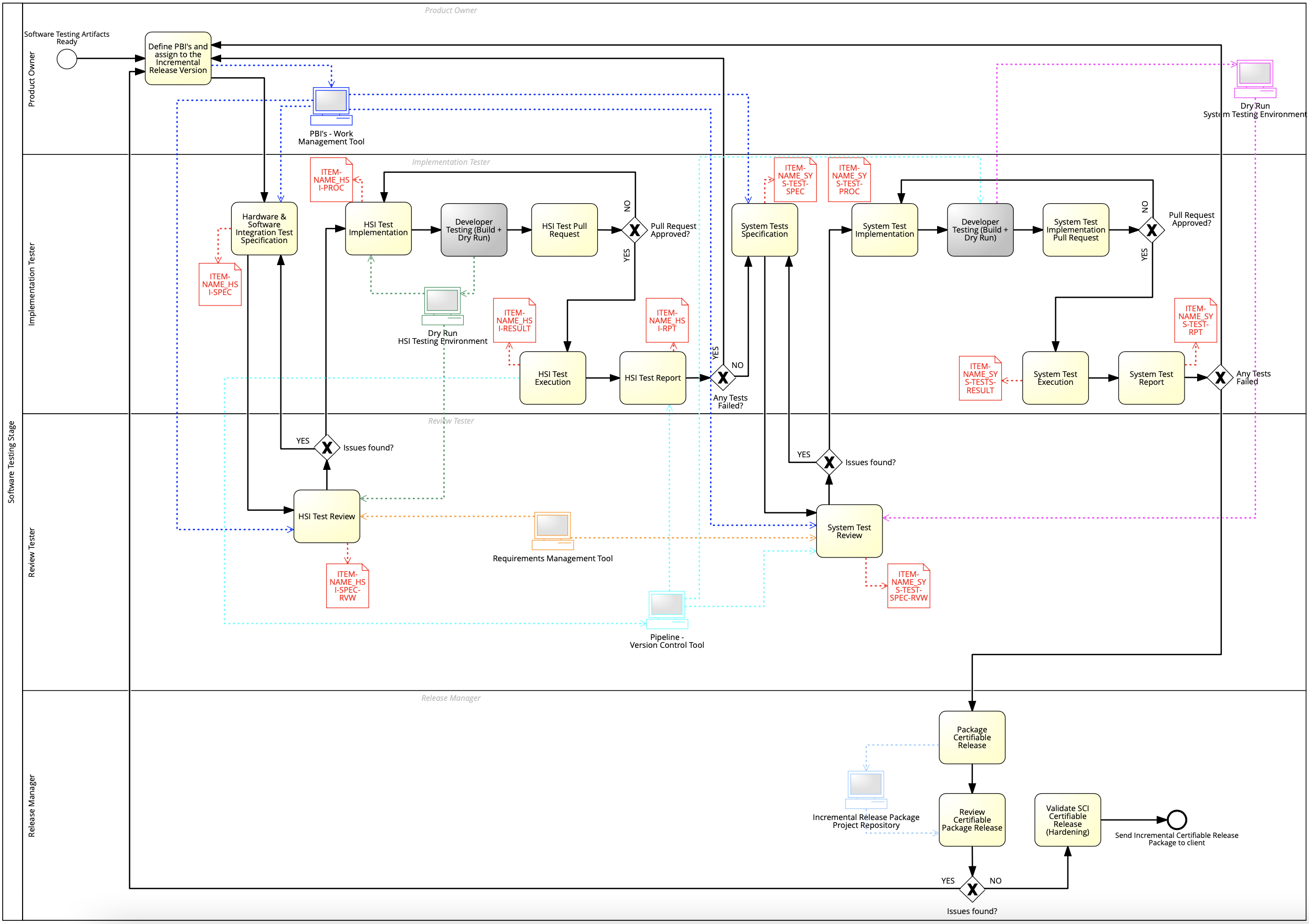}}
    \caption{Overview of the HSI \& System Testing Stage.}
    \label{bpmn-hsi-system-testing-stream}
\end{figure}

The key activities involved in the HSI \& System Testing Stage, include:
\begin{itemize}
    \item \textbf{Defining PBIs and assigning them to the Incremental Release Version}: define the PBIs considering the inputs from the previous stage.
    \item \textbf{Hardware \& Software Integration Test Specification}.
    \item \textbf{HSI Test Review}: review and generate the checklist.
    \item \textbf{HSI Test Implementation}: code the tests.
    \item \textbf{Developer Testing (Build + Dry Run)}: test the build and run HSI tests locally or in a simulation environment.
    \item \textbf{HSI Test Pull Request}: Creating the Pull Request in the VCS to trigger HSI test execution.
    \item \textbf{HSI Test Execution}: execute the tests.
    \item \textbf{HSI Test Report}: generate a report with the test results.
    \item \textbf{System Tests Specification}: Develop System tests.
    \item \textbf{System Test Review}: review system tests and generate the checklist.
    \item \textbf{System Test Implementation}: coding the system tests.
    \item \textbf{Developer Testing (Build + Dry Run)}: build and run sys tests locally or in a simulated environment.
    \item \textbf{System Test Implementation Pull Request}: trigger remote execution.
    \item \textbf{System Test Execution}: execute all System tests in a distinct environment.
    \item \textbf{System Test Report}: report  results.
    \item \textbf{Package Certifiable Release}: Generate the package with documentation of this and previous stages. This documentation shall be sent to the Certification Entity after the Hardening process.
    \item \textbf{Review Certifiable Release}: Manually review the package to ensure that all necessary files are included.
    \item \textbf{Validate SCI Certifiable Release (Hardening)}: perform a final ma\-nu\-al validation before sending the package and SCI to the client, including attaching all the documentation generated in each stage.
\end{itemize}

At the end of this stage, a new version of the Data Package is automatically generated after the Hardening step. This makes it different from the previous three stages, as the artifacts produced in this stage are automatically placed in the Data Package by the CMT and undergo a final manual validation. However, now the CPT is manually triggered to generate a set of certifiable documents because it contains all the required outputs according to the \emph{DO-178C} standard.

After a final manual review, this output can then be sent to the certification authorities (EASA in Europe).

\section{Threats to Validity}
\label{ThreatstoValidity}
Some limitations and threats to validity were identified during this study. First, the proposed mechanisms were developed during three years in close collaboration with aerospace industry experts working in a specific company, thus the proposed mechanisms are aligned with their specific practices. This means that our current solution may need to be tailored to the different practices (and toolsets) of other companies. 
Another major constraint is that to address the limitations of the Scrum Framework to accommodate safety-critical requirements, the authors designed the Scrum4DO178C model to incorporate additional V\&V steps and quality gates to ensure that the software meets the requirements of \emph{DO-178C}. This may require development teams to change their current practices, which can lead to cultural resistance.



\section{Conclusion}
\label{Conclusion}
We shall now revisit the research questions to summarize our findings.

\textbf{RQ1:Can automation mechanisms for documentation-related processes enhance the software certification process?} Yes, they can, spe\-ci\-fi\-ca\-lly for aerospace, our research focus. 
By automating the generation of necessary artifacts and connecting various tools used throughout the development process, these mechanisms drastically reduce manual effort and human errors, thus accelerating the overall certification timeline.

\textbf{RQ2:Which strategies can optimize a Document Management Tool (DMT) to make it \emph{quasi-autonomous}?} 1) Integrating existing tools (e.g., Jira, DOORS, Doxygen, BitBucket) to perform automatic updates and synchronization of documentation with ongoing development activities. 2) Using templates to create certification documents from sources and requirements, and 3) Compiling all relevant output, documentation and traceability data into a certifiable Data Package.


\textbf{RQ3:Is it feasible for a DMT to generate certifiable documentation automatically?} Yes, it is, as we were capable of automatically create documentation, such as SRD and SCI, from inputs extracted directly from the project repository. These were chosen because they are the most challenging artefacts to create. The remaining are considered trivial.  


Our next steps include: (i) Finalize the tools to include the remaining do\-cu\-ments, beyond the SRD and SCI; (ii) Validating the solution with more use cases from different aerospace companies.


In conclusion, by adopting a set of documentation management tools and following the strategies described, the current Waterfallish software development in the aerospace industry can evolve into a fully \emph{DO-178C} compliant iterative model, that is, more Agile.

\begin{credits}
\subsubsection{\ackname} The authors thank Critical Software SA\footnote{https://criticalsoftware.com/en} for granting access to real industry aerospace projects data used in this research; ARROW ECS Portugal\footnote{https://www.arrow.com/globalecs/pt/} and IBM\footnote{https://www.ibm.com/us-en} through António Lima, for facilitating access to the IBM DOORS industry-specific tool. Special thanks to Vitor Conceição, Sergio Santos, and Sergio Cardoso for their expertise in \emph{DO-178C} and design contributions, which significantly enhanced the success of the study. Furthermore, the authors appreciate the MSE\footnote{https://www.uc.pt/en/fctuc/dei/education/masters/master-in-software-engineering/} students for their collaboration, commitment, and contributions to this research.

\subsubsection{\discintname}
The authors have no competing interests relevant to this article's content to declare.
\end{credits}
%
%
%

\begin{thebibliography}{}

\bibitem{Ribeiro2023} Eduardo Ferreira Ribeiro, J., Silva, J. \& Aguiar, A. Weaving Agility in Safety-Critical Software Development for Aerospace: From Concerns to Opportunities. {\em IEEE Access}. \textbf{12} pp. 52778-52802 (2024)

\bibitem{RE_ribeiro2023beyond}Ribeiro, J., Silva, J. \& Aguiar, A. Beyond Tradition: Evaluating Agile feasibility in DO-178C for Aerospace Software Development.  (2023), https://arxiv.org/abs/2311.04344

\bibitem{RE_DO178C}SC-205, R. DO-178C - Software Considerations in Airborne Systems and Equipment Certification. (RTCA,2011)

\bibitem{RE_Rierson2013}Rierson, L. Developing Safety-Critical Software: A Practical Guide for Aviation Software and DO-178C Compliance. (CRC Press,2013)

\bibitem{RE_Marques2013}Marques, J., Yelisetty, S., Da Cunha, A. \& Dias, L. CARD-RM: a reference model for airborne software. {\em 2013 10th International Conference On Information Technology: New Generations}. pp. 273-279 (2013)

\bibitem{RE_SilvaCardoso2022}Rodrigues, J., Ribeiro, J. \& Aguiar, A. Improving Documentation Agility in Safety-Critical Software Systems Development For Aerospace. {\em 2022 IEEE International Symposium On Software Reliability Engineering Workshops (ISSREW)}. pp. 222-229 (2022)

\bibitem{zelkowitz_advances_2004}Zelkowitz, M. Advances in Computers: Advances in Software Engineering. (Elsevier,2004,7)

\bibitem{tordrup_meshing_2020}Tordrup, L. \& Nielsen, P. Meshing agile and plan-driven development in safety-critical software: a case study. {\em Empirical Software Engineering}. \textbf{25} pp. 1-28 (2020,3)

\bibitem{RE_Schwaber2020}Schwaber, K. \& Sutherland, J. Scrum Guide V7.  (2020)

\bibitem{RE_DO330}SC-205, R. DO-330 - Software Tool Qualification Considerations. (RTCA,2011)

\bibitem{ahmadi_achachlouei_document_2021}Ahmadi Achachlouei, M., Patil, O., Joshi, T. \& Nair, V. Document Automation Architectures and Technologies: A Survey. (Corporate Model Risk,2021,9), https://ui.adsabs.harvard.edu/abs/2021arXiv210911603A, Publication Title: arXiv e-prints ADS Bibcode: 2021arXiv210911603A Type: article

\bibitem{comoretto_documentation_2020}Comoretto, G., Guy, L., O'Mullane, W., Bechtol, K., Carlin, J., Klaveren, B., Roberts, A. \& Sick, J. Documentation automation for the verification and validation of Rubin Observatory software. {\em Modeling, Systems Engineering, And Project Management For Astronomy IX}. pp. 16 (2020,12), https://www.spiedigitallibrary.org/conference-proceedings-of-spie/11450/2561604/Documentation-automation-for-the-verification-and-validation-of-Rubin-Observatory/10.1117/12.2561604.full

\bibitem{lee_introduction_2012}Lee, J., Yang, S., Choi, J., Cheon, Y. \& Yun, J. Introduction to Automatic Generation of Design Documents for Flight Software using Doxygen. {\em Proceedings Of The Korea Information Processing Society Conference}. pp. 844-847 (2012), https://www.koreascience.or.kr/article/CFKO201221868477394.page, Publisher: Korea Information Processing Society

\bibitem{xsdoc}Aguiar, A. \& David, G. WikiWiki weaving heterogeneous software artifacts. {\em Proceedings of the 2005 international symposium on Wikis (WikiSym '05)}. Association for Computing Machinery, New York, NY, USA, pp. 67–74, https://doi.org/10.1145/1104973.1104980

\bibitem{lankester_implementing_2018}Lankester, R. Implementing Document Automation: Benefits and Considerations for the Knowledge Professional. {\em Legal Information Management}. \textbf{18}, 93-97 (2018,6), Publisher: Cambridge University Press

\bibitem{youn_software_2015}Youn, W., Hong, S., Oh, K. \& Ahn, O. Software Certification of Safety-Critical Avionic Systems: DO-178C and Its Impacts. {\em Aerospace And Electronic Systems Magazine, IEEE}. \textbf{30} pp. 4-13 (2015,4)

\bibitem{hilderman_understanding_2014}Hilderman, V. Understanding DO-178C Software Certification: Benefits Versus Costs. {\em 2014 IEEE International Symposium On Software Reliability Engineering Workshops}. pp. 114-114 (2014,11)

\bibitem{kennedy}Kennedy, J. \& Towhidnejad, M. Innovation and certification in aviation software. {\em 2017 Integrated Communications, Navigation And Surveillance Conference (ICNS)}. pp. 3D3-1-3D3-15 (2017)

\bibitem{dmitriev_lean_2020}Dmitriev, K., Zafar, S., Schmiechen, K., Lai, Y., Saleab, M., Nagarajan, P., Dollinger, D., Hochstrasser, M., Holzapfel, F. \& Myschik, S. A Lean and Highly-automated Model-Based Software Development Process Based on DO-178C/DO-331. {\em 2020 AIAA/IEEE 39th Digital Avionics Systems Conference (DASC)}. pp. 1-10 (2020,10), ISSN: 2155-7209

\bibitem{Sutherland2011}Sutherland, J. \& Ph, D. Future of Scrum : Creating a Scrum Company with a Type C All-At- Once Scrum. {\em Cambridge Innovation Center}., 30 (2011)

\end{thebibliography}
%

\end{document}